\documentclass[conference]{IEEEtran}
\makeatletter
\def\ps@headings{%
\def\@oddhead{\mbox{}\scriptsize\rightmark \hfil \thepage}%
\def\@evenhead{\scriptsize\thepage \hfil \leftmark\mbox{}}%
\def\@oddfoot{}%
\def\@evenfoot{}}
\makeatother
\usepackage{url}
\usepackage{graphicx}
\usepackage{amsfonts}
\usepackage{subfig}
\usepackage{amsmath}
\newcommand{\enc}{\texttt{Enc}}

\newcommand{\trans}{\text{\texttt{Transform}}}

\begin{document}

\title{Stream on the Sky: Outsourcing Access Control Enforcement for Stream Data to the Cloud}
\author{Tien Tuan Anh Dinh, Anwitaman Datta
\\ School of Computer Engineering, Nanyang Technological University, Singapore
\\ \{ttadinh,anwitaman\}@ntu.edu.sg
}
\maketitle
\begin{abstract}
There is an increasing trend for businesses to migrate
their systems towards the cloud. Security concerns that arise when outsourcing data and
computation to the cloud include data confidentiality and privacy. Given that a tremendous amount of
data is being generated everyday from plethora of devices equipped with sensing capabilities,
we focus on the problem of access controls over live streams of data based on triggers or
sliding windows, which is a distinct and more challenging problem than access control over
archival data. Specifically, we investigate secure mechanisms for outsourcing access control
enforcement for stream data to the cloud. We devise a system that allows data owners to specify 
fine-grained policies associated with their data streams, then to encrypt the streams
 and relay them to the cloud for live processing and storage for future use. The
access control policies are enforced by the cloud, without the latter learning about the data,
while ensuring that unauthorized access is not feasible. To realize these ends, we employ a
novel cryptographic primitive, namely proxy-based attribute-based encryption, which not only
provides security but also allows the cloud to perform expensive computations on behalf of the
users. Our approach is holistic, in that these controls are integrated with an XML based
framework (XACML) for high-level management of policies.  Experiments with our prototype
demonstrate the feasibility of such mechanisms, and early evaluations suggest graceful
scalability with increasing numbers of policies, data
streams and users.  
\end{abstract}

\section{Introduction}
An enormous amount of data is being generated continuously, by organizations and individuals
carrying out their day-to-day activities, as well as by dedicated sensing and monitoring
infrastructures.  Examples include financial services for monitoring stock
prices~\cite{xignite}, sensor networks for meteorological, environmental~\cite{noaa}, battle
field and traffic control~\cite{abadi03,arasu04} monitoring. As personal devices, especially
those equipped with sensing capabilities, are enjoying an unprecedented growth, they are also
becoming a prominent source of stream data. Applications such as participatory
sensing~\cite{dutta09,lane12} and personal health monitoring~\cite{healthFrontier} collect
continuous data from sensors fitted in smart phones or in other mobile devices.


The abundance of such data brings many new opportunities, such as real time decision making
and resource management at various scales - from personal area or home networks to smart
planet. Often, these applications assume sharing and mash-up of data from multiple sources,
possibly created and owned by different stake holders. One important requirement is
to equip data owners with adequate control for determining different granularity in
which the sharing is done with various parties. Another requirement is the infrastructure over
which such sharing can be done in a scalable manner.

One can presume that the need for a scalable infrastructure can be readily realized thanks to
the advent of cloud computing. Businesses are moving their computing systems cloud-ward,
availing themselves of the elastic resources, ease of management and good cost-benefit
trade-off~\cite{armbrust09,hajjat10,chen11,tak11}. While recent advances in the technologies
have given users more control and better performance~\cite{ballani11,shieh11}, limited progress
has been made towards security guarantees, particularly vis-a-vis sharing data with multiple
other parties in possibly different granularity. In general, many
argue~\cite{armbrust09,popa11,popa11a} that security remains one of the biggest obstacles to be
overcome before the potential of the cloud can be fully realized.

This paper is an important step exploring the design space of outsourcing the enforcement of
access control of live data streams to the cloud. Our goal is to design a system that supports
fine-grained access control policies in which most of the expensive computations are done by
the cloud. We assume that a data owner outsources its stream data to an untrusted
cloud. The system must allow the data owner to specify fine-grained access control policies for
the data stream. It must ensure security with respect to access control enforcement, i.e. no
unauthorized access is allowed even if the cloud colludes with dishonest data users. Finally,
the system must protect privacy of the outsourced data.

Access control for stream data is more challenging than for traditional non-stream (archival)
data. In archival databases, access is defined on views~\cite{ramakrishnan02} which are
constructed by querying the databases. Since views are static and can be pre-computed, this
technique is not applicable to stream data because of the potentially infinite size and queries
being continuous and are carried out over newly arrived data. Most critically, the nature of
stream data demands support for more fine-grained policies, especially those involving temporal
constraints~\cite{carminati07}. Specifically, most policies can be categorized as
\emph{trigger} or \emph{sliding window} policies:

- Trigger: A user is given access to a data record when its content satisfy a certain
  condition. For example, in a personal health monitoring system~\cite{healthFrontier}, a user
tracks his blood pressure, heart rate, blood sugar, etc. using a portable device at regular
intervals. On the one hand, such data is crucial for doctors to monitor patient recovery and to
make early diagnosis. On the other hand, the user is concerned with his privacy and would not
wish to disclose all the measurements. As a result, he can specify a policy dictating that
certain doctors have access to his data only when his blood pressure exceeds a threshold.

- Sliding window: A user is given summaries of the data over specific windows, as opposed to
  having access to raw data. For example, a financial company monitoring stock prices every
$0.5$ second can sell its raw data for a very high price. It can also offer more
\emph{coarse-grained} packages at lower prices which contain only the average stock prices over
$s$-second interval ($s > 0.5$). This gives the data owner greater flexibility in managing and
generating profit from its data.

A naive design to realize these goals would be to encrypt the data using a symmetric encryption
scheme, store it on the cloud and distribute the decryption keys to authorized users. Such a
system~\cite{popa11a,kallahalla03} relies on the data owner for access control enforcement,
while the cloud acts only as a transit storage provider. Use of a more advanced encryption
scheme, namely attribute-based encryption (ABE) scheme~\cite{goyal06,bethencourt07}, can help
relieve the data owner from much of the key management tasks. However, ABE deals only with
individual data objects and therefore can not naturally support sliding window policies.
Additionally, ABE effectively pushes the access control management task towards the end-user
who needs to check all the ciphertexts and discard that that will result in invalid decryption.
Such a modus operandi is inefficient and does not scale for large systems with many different
data streams.

In this paper, we propose to use a proxy attribute-based encryption scheme~\cite{green11} for
trigger policies, and extend it with support for sliding window policies. Out system ensures
access control while also offloading expensive computations to the cloud. Furthermore, we
decouple the task of policy management from security enforcement by using the XACML
framework~\cite{xacml}. The cloud handles this task seamlessly, and as a result the end-user
only receives and decrypts ciphertexts of his authorized data.  Adding an explicit layer of
management on top of ABE helps the system scale better with more data streams and more
policies. In summary, our main contributes are as follows:

1. We propose an extension to an ABE scheme which provides support for sliding window access
control.

2. We design a system allowing data owner to outsource stream data and access control
enforcement to the cloud. The system supports both trigger and sliding window policies in a
secure manner. It is also scalable, in the sense that most of the expensive operations are done
by the cloud. We achieve both security and scalability by combining novel cryptographic
primitives with the standard XACML framework for policy management.

3. We implement a prototype of our system. Through preliminary evaluation, we find that the
overhead is reasonable.

The remaining of the paper is organized as follows. The next section
presents the system and security model. Section~\ref{sec:protocol}
details the cryptographic protocols. It is followed by the design
for policy management. We present our prototype and preliminary
evaluation in Section~\ref{sec:evaluation}. Related work is
discussed in Section~\ref{sec:relatedWork} before we conclude in
Section~\ref{sec:conclusions}.

\section{System Model}
\label{sec:model}
Our system consists of a number of \emph{\textbf{data owners}} (or owners),
one \emph{\textbf{cloud}} provider and a number of \emph{\textbf{data users}} (or users). As
depicted in Figure~\ref{fig:context}, a data owner generates a stream of data and sends it to
the cloud. Several users interested in the data will retrieve it through the cloud. The owner
and users agree on the access policy before-hand. In summary, the data is \emph{outsourced} to
the cloud where it will be stored, managed and distributed to a set of users.

\subsection{Data Model}
For simplicity, we assume that data generated by the owner are key-value tuples. A stream $D$
is a sequence of $(k_i,v_i$) for $i = 0,1,2,..$ where $k_i, v_i$ are integers and the value
$v_i$ belongs to a known, finite domain. In a participatory sensing application, the $k_i$ may
represent a geographical location while $v$ may represent a measurement of air pollution. In a
time-series application, such as weather monitoring, $k_i$ may represent a time-stamp value
while $v_i$ a temperature reading.

For a time-series data stream, i.e. $D = \langle (0,v_0), (1,v_1), (2,v_2), ..\rangle$, we
define a \emph{data window}, denoted as $w(s,e)$, as a sequence of data values whose keys are
in between a starting index $s$ and an ending index $e$. That is: $ w(s,e) = \langle\, v_i
\,|\, s \leq i < e \rangle$.  A \emph{non-overlapping sliding window} --- denoted as
$sw(\alpha,\beta)$ --- is a sequence of non-overlapping data windows starting from index $\alpha$, having size
$\beta$. More precisely, the $i^{th}$ sliding window is
$sw(\alpha,\beta)[i] = w(\alpha*i,\alpha*i+\beta)$.

\subsection{Access Control Model.}
An access control policy is defined by the data that an authorized user can read. We
consider two types of policies: trigger and sliding window. In the
former, the user is granted access to a tuple $(k_i,v_i)$ (i.e. it can read the value $v_i$)
if $k_i$ is equal to or exceeds a threshold $\theta$. Denote $\tau(\theta,op)$ as the
trigger policy, where $\theta$ is a threshold value and $op$ is a comparison operator,
then:
\[
\tau(\theta,op) = \langle\, v_i \,|\, k_i\, op\, \theta = true \,\rangle
\]

\begin{figure}
\centering
\includegraphics[scale=0.35]{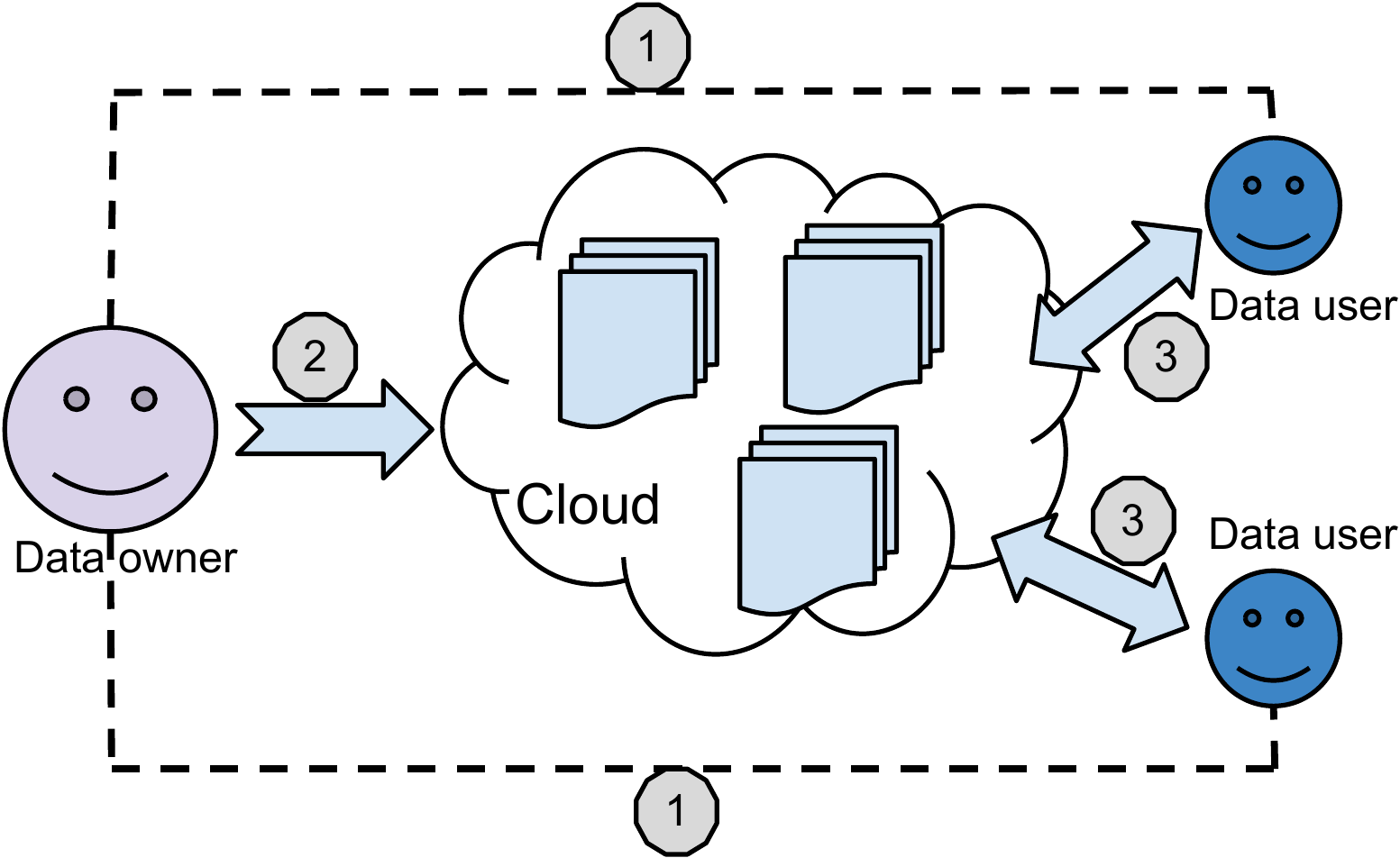}
\caption{The context: (1) \emph{negotiation phase}: the owner and the
user agree on an access control policy (2) \emph{outsourcing
phase}: the owner encrypts its data and forwards to the cloud (3)
\emph{relaying phase}: the cloud processes and forwards the data to
the authorized users.} \label{fig:context}
\end{figure}

In a sliding window policy, an user can only access the averages of the data tuples that
constitute a non-overlapping sliding window. In other words, the user cannot read individual
data tuples, but only the summaries. Denote $\pi(\alpha,\beta)$ as the sliding window policy with
$\alpha$ being the starting index and $\beta$ being the window size, we have: \[
\pi(\alpha,\beta) = \langle \frac{\sum v}{\beta}\, | \, v \ \text{in} \ sw(\alpha,\beta)[i]; i =
1,2,.. \rangle
\]
As an example, suppose a data owner is to share a weather database containing temperature
readings collected at five-minute intervals starting from time $0$. A sliding window policy may
require that the user can only access the  average temperature
every $1$ hour, starting from time $3$. This corresponding to the policy $\pi(3,12)$, meaning that the
user can only read the averages of the windows $[3-14]$, $[15-26]$, $[27-38]$, etc.


\subsection{Trust Model}
We assume that data owners are honest. They outsource data streams to the cloud and wish to
protect privacy of their data against the cloud. They define access policies for a set of users
and wish to enforce the policies in a secure manner that does not permit unauthorized data
access. The cloud is not trusted in that it tries to learn sensitive information in the data, and may
collude with rogue users in order to provide them unauthorized access to the data. However, the
cloud is trusted to carry out computations delegated to it by the owners and users. This means
it does not skip or distort computations. A data user may be dishonest, in which case he
attempts to gain unauthorized access, for which he may collude with the cloud. Finally, we
assume that users do not collude with each other (we explain the reason for this assumption
shortly).

\subsection{System Design Goals}
Given the models above, we now state informally the goals that our system aims to achieve.

\textbf{Goal 1 (privacy).}  The cloud cannot learn the outsourced data in plaintext.

\textbf{Goal 2 (access control, trigger).} An user given access according to the policy $\tau(\theta,op)$
cannot access values belonging to another policy $\tau(\theta',op')$ where $\theta \neq \theta'$
or $op \neq op'$.

\textbf{Goal 3 (access control, sliding window).} An user given access according to the policy $\pi(\alpha,\beta)$
cannot access values belonging to any different policy. Specifically, the user cannot see the
values belonging to $\pi(\alpha',\beta')$ where $\alpha'<\alpha$ or $\alpha' \neq \alpha \ mod
\ \beta$. This means the user does not have access to data windows starting before $\alpha$.
He can access windows starting from later indices, but those indices must be a multiple of
$\alpha$, which is the same as skipping a number of windows. Moreover, the user cannot see the
values defined by $\pi(\alpha',\beta')$ where $\beta'<\beta$ or $\beta' \neq 0 \ mod \ \beta$.
This means the user cannot access data at finer granularity, nor can it
access windows whose sizes are not multiples of $\beta$. For instance, if an user is given
access to windows of size $5$, he must not be able to access windows of size $3$ or
size $7$. It is worth noticing that the assumption about users not colluding with each others
is necessary to achieve this goal. If two users with $\pi(\alpha,\beta)$ and
$\pi(\alpha,\beta')$ where $\beta \neq 0 \ \text{mod} \ \beta'$ collude, they can derive
windows of size $|\beta-\beta'|$ or even windows of size $1$.

\textbf{Goal 4 (scalability).} Most of the expensive computation should be off-loaded to the
cloud. The protocols involving the data owners and users should be light-weight.

\section{The Protocols}
\label{sec:protocol}
\subsection{Overview}
Our protocols for both trigger and sliding window policies consist of 3 phases, as illustrated
in Figure~\ref{fig:context}. First, the data owner and user negotiate an access control
policy which can be either a trigger or sliding
window policy (negotiation phase). Next, the data owner encrypts its data and sends
it to the cloud (outsourcing phase). Upon receiving
the ciphertexts, the cloud performs transformation on the ciphertexts and forwards the results
to the authorized users (relaying phase). The users receive messages from the cloud and decrypt them to obtain
the plaintext data.

A naive approach would be for the data owner to generate one encryption key for each policy,
share this key to the authorized users, and encrypt the data using this key. This does not
scale for two reasons. First, the amount of duplicated data that must be encrypted grows
linearly with the number of policies. Second, for a sliding window
policy $\pi(\alpha,\beta)$, the owner has to transform the data locally and update the
encrypted version to the cloud. This process must be done for every combination of $\alpha$ and
$\beta$.

In our protocol, there is only encrypted copy of the data stored at the cloud for the
trigger policies, and up-to $w$ copies for sliding window policies where $w$ is the number of
unique window sizes. To this end, we make use of three important cryptographic primitives: attribute-based
encryption (ABE), proxy re-encryption (PRE) and additive homomorphic encryption (AHE). ABE is
used to enforce threshold conditions specified in trigger policies, as well as the
condition on the starting index in sliding window policies. PRE is used by the cloud to
transform ABE ciphertexts to simpler ciphertexts that can be decrypted by users. The aims of
this transformation is to relieve the users from carrying out expensive operations involving
ABE. For sliding window policies, AHE enables the cloud to compute the encrypted aggregates of
the data windows over the encrypted text without breaking data privacy.

\subsection{Cryptographic Primitives}
\subsubsection{Attribute Based Encryption.}
Attribute-Based Encryption (ABE) schemes produce ciphertexts that
can only be decrypted if a set of attributes satisfies certain
conditions (or policies). Two types of ABE exists: key-policy
ABE~\cite{goyal06} (KP-ABE) and ciphertext-policy
ABE~\cite{bethencourt07} (CP-ABE). In both cases, a policy is
expressed as an access structure over a set of attributes. For
example, if the policy is defined as the predicate $T = a_1 \,
\wedge \, (a_2 \vee a_3)$, then $T(\{a_1\}) = T(\{a_2,a_3\}) =
\mbox{false}$, $T(\{a_1,a_2\}) = T(\{a_1,a_3\}) = \mbox{true}$. The
plaintext can be recovered only if evaluating the policy over the
given attributes returns true.

In KP-ABE, the message $m$ is encrypted with a set of attributes $A$
and an access structure $T$ is embedded in a decryption key. In
CP-ABE, $T$ is embedded within the ciphertext and the user is given
a set of attribute $A$. In both cases, decryption will succeed and
return $m$ if $T(A) = \mbox{true}$. KP-ABE and CP-ABE are based on
secret sharing schemes that generate shares in a way that the
original secret will only be reconstructed if a certain access
structure is satisfied. Constructions of KP-ABE and CP-ABE for any
linear secret sharing scheme (modelled as a span program) are
described in~\cite{goyal06,bethencourt07}. In practice, the schemes
rely on threshold tree structures for secret sharing, in which
non-leaf nodes are threshold gates (AND and OR).

We adopt KP-ABE in our work, which is more suitable than CP-ABE,
because the attributes are based on the plaintext message (the key
value $k$). Although KP-ABE and CP-ABE are similar and in some cases
can be used interchangeably, the former is more data-centric
(concerning the question of who gets access to the given data),
whereas the latter is more user-centric (concerning the question of
which data the given user has access to). For the trigger policy
$\tau(\theta,op)$ we encrypt $(k_i,v_i)$ using KP-ABE with $k_i$ as
the attribute and $k \ op \ \theta = \mbox{true}$ as the policy.
Similarly, for the sliding window policy $\pi(\alpha,\beta)$, we
encrypt $(k_i,v_i)$ using $k_i$ as the attribute and $k \geq \alpha$
as the policy.

Since the policies involve arithmetic comparison, we make use of the \emph{bag-of-bits} method
that translates an integer into a set of attributes. For example, with $64$-bit integers,
an attribute $3$ can be mapped to $\{x..x1x, x..x1\}$, and the policy $k \geq 11$ can be mapped
into $(ge2exp4 \,\vee \, ge2exp8 \,\vee\, ge2exp16 \,\vee\, ge2exp32) \vee (x..1xxx \,\wedge\,
(x..1xx \, \vee \, (x..x1x \,\wedge\, x..xx1)))$ where $ge2expm$ is an attribute representing
values greater than or equal to $2^m$. It can be seen that a policy involving equality
condition often involves more attributes than a policy with an inequality condition.

\subsubsection{Proxy Re-Encryption.}
A Proxy Re-Encryption scheme (PRE)~\cite{blaze98} enables a third-party to transform one
ciphertext to another without learning the plaintext. Using a \emph{transformation key} $K_{U
\to V}$ it can convert $c = \enc(K_U,m)$ to $c' = \enc(K_V,m)$ which can be decrypted with
$K_V$, without learning $m$. Traditional PRE schemes have been used for distributed file
systems, in a way that saves the data owner from having different copies of the same data for
different users. In this work, we extend a recent proxy-based ABE scheme~\cite{green11} which
can transform a KP-ABE ciphertext into a Elgamal-like ciphertext whenever the access structure
is satisfied. This scheme has the benefit of traditional PRE schemes, and it relieves users
from expensive computations involving ABE.

\subsubsection{Additive Homomorphic Encryption.}
Sliding window policies require that authorized users have access to the averages of data values
within non-overlapping windows. To allow for such an aggregate operation over
ciphertexts, the encryption scheme must be additively homomorphic, i.e.  $\enc(K,m_1) \oplus
\enc(K,m_2) = \enc(K,m_1+m_2)$ for an operator $\oplus$. Paillier~\cite{paillier99}
is one of such schemes, which is based on composite residuosity classes of group
$\mathbb{Z}_{N^2}$ where $N$ is at least 1024-bit.

In this work, we use an Elgamal-like encryption scheme to achieve
additive homomorphism, which can also be integrated with ABE.
Essentially, consider a multiplicative group $\mathbb{Z}_p$ with generator
$g$. For $m \in \mathbb{Z}_p$, $K$ and $r$ as random values of
$\mathbb{Z}_p$, then $\enc(K,m) = (g^r, g^m.g^{r.K})$. The
homomorphic property holds, because $\enc(K,m_1).\enc(K,m_2) =
\enc(K,m_1+m_2)$. The drawback of this scheme is that decryption
requires finding the discrete logarithm of $g^m$ in base $g$. As in
other works using a similar scheme~\cite{cramer97,zhong07,ugus09},
we assume that the plaintext domain is finite and small so that
discrete logarithms can be computed by brute-force~\cite{pollard78}
or even pre-computed and cached. For many types of applications
where data values belong to a small set (such that temperature
values), this is a reasonable assumption. Note that the security is
not compromised due to the mapping of $g^m$ to $m$, which merely acts as an
additional layer used to achieve homomorphism.

To prevent a user from learning the individual values constituting a
window, we introduce a blind factor into the ciphertext, which can
only be removed when a proper number of ciphertexts are put
together. Suppose the windows are of size $2$, the owner generates
$\delta_1,\delta_2$ randomly and give $\sigma = \delta_1+\delta_2$
to the user. The ciphertext $c_1 = (g^{r_1},g^m.g^{r_1+\delta_1+K})$
and $c_2 = (g^{r_2},g^{m_2}.g^{r_2+\delta_2+K})$ cannot be decrypted
individually, since $\delta_1$ and $\delta_2$ are unknown to the
user. However, $c_1.c_2 = (X = g^{r_1+r_2}, Y =
g^{m_1+m_2}.g^{K(r_1+r_2+\delta_1+\delta_2)})$ can be decrypted by
computing $g^{m_1+m_2} = \frac{Y}{X^K.g^{\sigma.K}}$ and recovering
$m_1+m_2$ by taking its discrete logarithm. In our protocol, the
user only needs to know $\sigma$ for the starting window, with which
the unblind factors for the subsequent windows can be efficiently
computed.

\subsection{Protocol for Sliding Window Policies}
\begin{figure}
    \centering
    \includegraphics[scale=0.45]{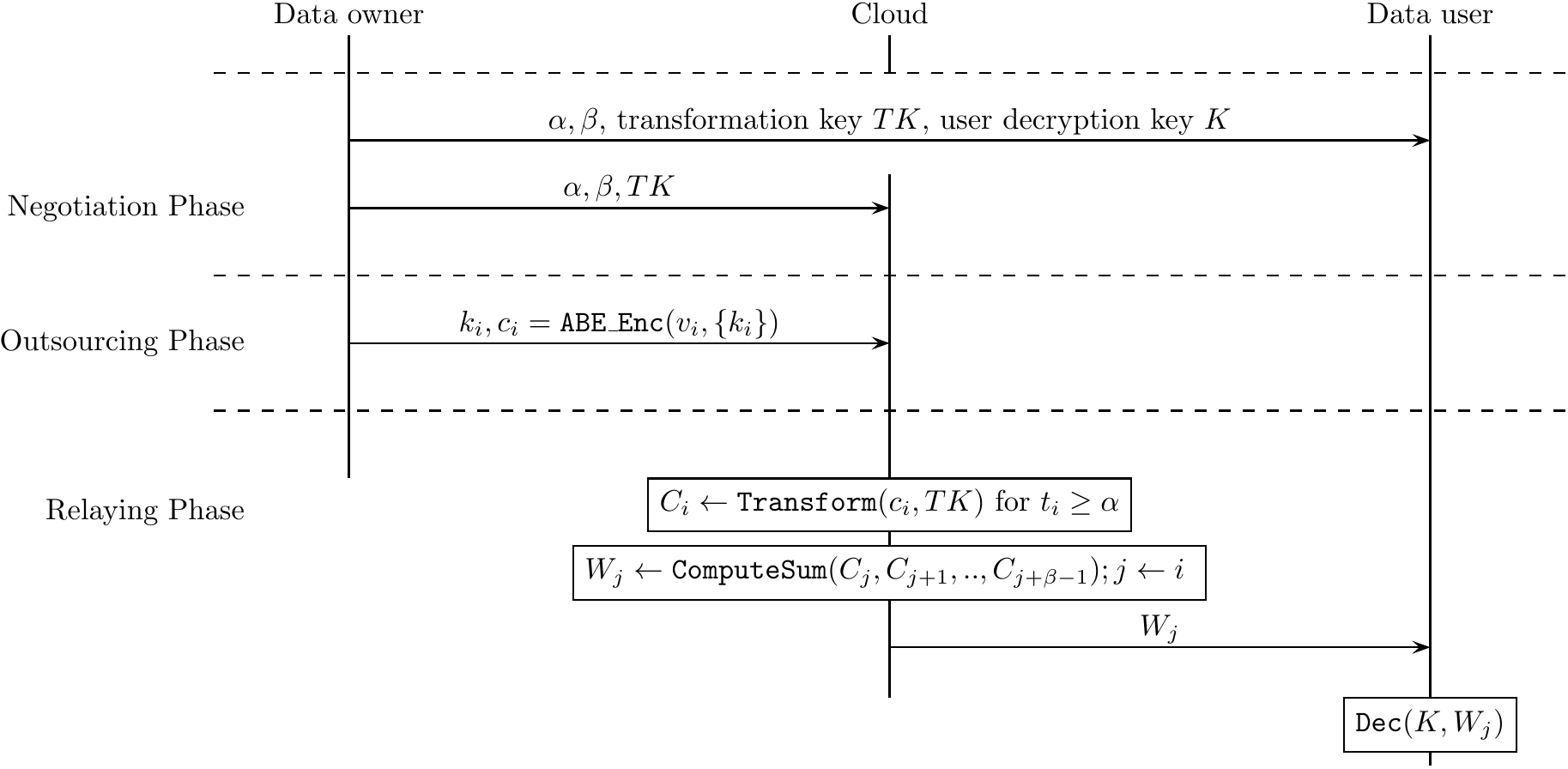}
    \caption{High-level protocol}
    \label{fig:highLevelProt}
\end{figure}
Figure~\ref{fig:highLevelProt} illustrates the high-level protocol for sliding window policies.
In the negotiation phase, data owner and user agrees on the sliding window parameters
$\alpha,\beta$. The owner generates an access tree $T$ for the condition $k \geq \alpha$, a
transformation key $TK$ and a decryption key $K$ for the user. It sends $(\alpha,\beta,TK,K)$
to the user and $(\alpha, \beta,TK)$ to the cloud. For every tuple $(k_i,v_i)$, the owner
encrypts $v_i$ using KP-ABE with attribute $k_i$.  Each tuple is encrypted $\omega$ times, each
for a different window size $\beta$ and using a different blind factor. During the relaying
phase, the cloud selects ciphertexts whose attribute is greater or equal to $\alpha$,
and transforms them using $TK$ into other ciphertexts that can be decrypted with another key
$K'$ $(K \neq K')$. Every $\beta$ of the transformed ciphertexts are used as inputs to the
\texttt{ComputeSum} function which produces an encrypted sum value which can be decrypted
using $K$.

Our protocol relies on the properties of \emph{bilinear maps} for security. Let $G,G_1$
be two multiplicative groups of prime order $p$, $g$ be the generator of $G$. $e:G
\times G \to G_1$ is a bilinear map if it is efficiently computable, $e(u^a,v^b) =
e(u,v)^{a.b}$ for $u,v \in G$ and   $e(g,g) \neq 1$. Given random group elements $g^a,
g^b, g^c$, it is difficult to compute $e(g,g)^{abc}$ (Computational Bilinear Diffie-Hellman
Assumption). Also, it is difficult to distinguish $e(g,g)^{abc}$ from a random element
$e(g,g)^z$ (Decisional Bilinear Diffie-Hellman Assumption).

\textbf{Negotiation phase.}
During setup, all parties agree on a security parameter $\lambda$.  Let $U$ be the attribute
universe, i.e. $U = \{x..x0, x..x1, x..0x, ...\}$. The owner chooses a master secret value $y
\in \mathbb{Z}_p$, and other secret $t_1, t_2,.. t_{|U|}$. For every window size $\beta$, it
generates a set of random values $R = \{r_0, r_1,..,r_{\beta-1}\}$. The public key contains
$e(g,g)^y, g^{t_1}, g^{t_2}, .., g^{t_{|U|}}$.

Let $u$ be an user with a sliding policy $\pi(\alpha,\beta)$. The owner generates a
random value $z_u \in \mathbb{Z}_p$. Next, the policy $k \geq \alpha$ is converted into an
access tree whose leaves are the bag-of-bit representation of $k$, as explained
in~\cite{goyal06}. Every non-leaf node is a threshold gate: an AND node is a 2-out-of-2 threshold
gate, an OR node is a 1-out-of-2 gate. This tree is used to share the master secret $y$. For a
node $x$ with threshold $d_x$, choose a random polynomial $q_x$ of degree $d_x-1$ (so
that $x_d$ points are needed to reconstruct the polynomial). For the root node, set $q_r(0) =
y$. For any other node, set $q_x(0) = q_{parent(x)}(index(x))$ where $q_{parent(x)}$ is the
polynomial of $x$'s parent and $index(x)$ is the node index. The leaf node $x$
associated with an attribute $i$ is given the value $D_x = g^{\frac{q_x(0)}{z_u.t_i}}$.

For the policy $\pi(\alpha,\beta)$, the owner computes:
\[
\sigma(\alpha,\beta) =  \sum_{j=0}^{\beta-1} 2^{\lceil (\alpha+j)/\beta \rceil}R[\alpha+j \ mod \ \beta]
\]
Finally, the user is sent
\[(\alpha, \beta, \{D_x\}, (z_u, \sigma(\alpha,\beta)))
\]
where $\{D_x\}$ and $(z_u,\sigma(\alpha,\beta))$ are the
transformation and user decryption keys respectively.

\textbf{Outsourcing phase.}
To encrypt $(k,v) \in D$, the data owner first translates $k$ into bag-of-bit attributes $B_k$.
Next, it chooses a random value $s_k \in \mathbb{Z}_p$ and computes $E'_i = g^{t_i.s_k}$ for all
attribute $i \in B_k$.  It also computes:
\[
E(k,v) = g^{v+2^{\lceil k/\beta \rceil}.R[k \ mod \ \beta]}.e(g,g)^{y.s_k}
\]
Finally, the following ciphertext is forwarded to the cloud:
\[
c_k = (k, B_k, E(k,v), \{E'_i\})
\]

\textbf{Relaying phase.}
The transformation proceeds recursively as follows. When $x$ is a leaf node associated with attribute $i$:
\[
    \trans(x) = \left\{
                \begin{array}{l}
                e(D_x,E'_i) = e(g,g)^{\frac{s_k.q_x(0)}{z_u}}\ \text{when $i \in B_k$}\\
                \bot \qquad \text{otherwise}
                \end{array}
                \right.
\]

When $x$ is a non-leaf node, we call $\trans(w)$ for all $w$ that are $x$'s children.
Let $\Delta_{i,S}(x) = \prod_{j \in S, j\neq i}\frac{x-j}{i-j}$ be the Lagrange
coefficient for an element $i \in \mathbb{Z}_p$ and $S \subseteq \mathbb{Z}_p$.
Let $F_w$ be the output of $\trans(w)$, and $S_x$ be the set of $x$'s children such
that $F_w \neq \bot$. If $S_x = \emptyset$, return $\bot$. Otherwise:
\begin{align*}
    F_x & = \trans(x)\\
    & = \prod_{w \in S_x}F_w^{\Delta_{index(w),S'_x}(0)} \ \text{where
$S'_x=\{index(w) \,|\, w \in S_x\}$}\\
    & = e(g,g)^{s_k.q_x(0)/z_u}
\end{align*}
By calling $\trans$ on the root node of the access tree, we obtain $\trans(root_k) =
e(g,g)^{y.s_k/z_u}$. Let $C_k = (E(k,v), \trans(root_k))$. For the $i^{th}$ window, the
cloud has to gather the full set of $\mathbb{C} = \{C_k, C_{k+1}, .., C_{k+\beta-1}\}$ for $k = \alpha+i.\beta$. Then, it
executes \texttt{ComputeSum} as follows:
\begin{align*}
(E_1, E_2) &= \texttt{ComputeSum}(\mathbb{C}) = \prod_{j=0}^{\beta-1} C_{k+j}\\
&= (\prod_{j=0}^{\beta-1} E(k+j,
v_{k+j}), \prod_{j=0}^{\beta-1} \trans(root_{k+j}))
\end{align*}
and sends $(i = \frac{k-\alpha}{\beta}, E_1, E_2)$ to the user.

Having received this message from the cloud, the user computes:
\begin{align*}
w_i &= \frac{E_1}{E_2^{z_u}} =
\frac{E(k,v).E(k+1,v_1)..E(k+\beta-1,v_{\beta-1})}{(e(g,g)^{y.s_k/z_u}..e(g,g)^{y.s_{k+\beta-1}/z_u})^{z_u}}\\
&=\frac{g^{(v_k+..+v_{k+\beta-1})+2^i.\sigma(\alpha,\beta)}.e(g,g)^{(s_k+..+s_{k+\beta-1}).y/z_u}}{(e(g,g)^{(s_k+..+s_{k+\beta-1}).y/z_u})^{z_u}}\\
&=g^{(v_k+..+v_{k+\beta-1})+2^i.\sigma(\alpha,\beta)}
\end{align*}
Finally, the average is computed as:
\[
avg_i = \frac{\mbox{discreteLog}(\frac{w_i}{g^{2^i.\sigma(\alpha,\beta)}})}{\beta} =
\frac{v_k+..+v_{k+\beta-1}}{\beta}
\]

\vspace{0.3cm}

\subsection{Protocol for Trigger Policy}
The protocol for trigger policies is very similar to that for sliding window. Its main
differences are that the access policies  may involve other conditions besides  $\geq$ as in
sliding window, and that no blind factor is needed since authorized users can access individual
data tuples.

During the negotiation phase, the owner create the public key containing $e(g,g)^y,
g^{t_1}, g^{t_2},.., g^{t_{|U|}}$ as before. Next, the access structure for the policy $k \, op
\, \theta$ is constructed. The
transformation key contains $\{D_x\}$ for all leaf node $x$, and the user decryption key is
$z_u$. During the outsourcing phase, a ciphertext sent to the cloud is generated in the same
way as with sliding window policy, except for $E(k,v) = g^v.e(g,g)^{y.s_k}$. During
transformation, the cloud computes $\trans(root_k) = e(g,g)^{y.s_k/z_u}$ as before and
sends $C_k = (E(k,v), \trans(root_k))$ to the user, who decrypts it by taking the discrete
logarithm of
$g^v = \frac{E(k,v)}{(\trans(root_k))^{z_u}}$.

\vspace{0.3cm}

\subsection{Discussions}
\textbf{Security.}
The  protocols described above provide data privacy with respect to the cloud, because we
use an encryption scheme which is an extension of the proxy-based ABE scheme proposed
in~\cite{green11}.  As shown in \cite{green11}, the scheme is Replayable-CCA (or RCCA) secure.
Our encryption is secure in the standard model (as opposed to security in random oracle model),
since we use the small-universe construction of ABE.

The access control property for trigger policies hold as a result of
ABE. For sliding window policies, we now discuss an informal proof
showing that unauthorized access is not possible. Particularly, we
need to show that an user given access according to a policy
$\pi(\alpha,\beta)$ cannot decrypt values belonging to other
policies $\pi(\alpha',\beta')$ where $\alpha' < \alpha$, $\beta' <
\beta$, $\alpha' \neq \alpha \ mod \ \beta$ or $\beta' \neq 0 \ mod
\ \beta$. Since the ciphertext is encrypted with the attribute $k$
and the policy embedded in the transformation key requires $k \geq
\alpha$, the $\trans$ operation will fail for $k < \alpha$. Thus,
the user is prevented from learning the values in
$\pi(\alpha',\beta')$ for $\alpha'<\alpha$. From
$\sigma(\alpha,\beta)$, one can infer the value
$\sigma(\alpha',\beta)$ for $\alpha'>\alpha$ and $\alpha' = \alpha \
mod \ \beta$, since $\sigma(\alpha',\beta) =
2^{\alpha'/\alpha}.\sigma(\alpha,\beta)$. However, for $\alpha' \neq
\alpha \ mod \ \beta$, deriving $\sigma(\alpha', \beta')$ from
$\sigma(\alpha,\beta)$ is not possible as it requires knowing the
sum of a subset of $R$. Therefore, the user cannot recover the
plaintext because the unblind factors are incorrect. This similarly
holds when $\beta' < \beta$ or $\beta' \neq 0 \ mod \ \beta$,
because computing the blind factors $\sigma(\alpha',\beta')$ in
these cases requires knowing individual values of $R$.

\textbf{Scalability.} The most expensive cryptographic operations
are exponentiations and pairings. The latter are done only by the
cloud during transformation. The number of pairings is proportional
to the size of the access tree.  Notice that the access tree is
likely to be smaller for a sliding window policy than for a trigger
policy, as there is a smaller number of attributes involved in
constructing $k > x$ than in constructing $k = x$. Without the
cloud, the user will have to perform these pairings themselves.
Instead, in our protocol, the user only performs simple operations
such as inversion, multiplication, and only one exponentiation.
Having to find discrete logarithm during decryption is a potential
bottleneck. But as discussed earlier, we assume the data value
domain is known and finite, hence the user may pre-compute the
discrete logarithm, thus reducing this problem into a simple table
look-up. If $m$ is the maximum value that $v$ can take, the look-up
table has the size of $m$ for trigger policies, and $b.m$ for
sliding policies where $b$ is the maximum window size.

The data owner's computations are during the negotiation phase and
the outsourcing phase. Public and master key are generated only
once, hence the cost will amortize over time. Generation of
transformation keys is done once per policy. Compared to data
encryption during the outsourcing phase, this operation is much less
frequent and can be considered as a constant. Thus to say, the
overall setup cost is a one-off cost which amortizes over time.
During the outsourcing phase, the owner performs one encryption per
data value for trigger policies, and $w$ encryptions for sliding
window policies ($w$ being the number of window sizes). The cost per
encryption is roughly the same for both trigger and sliding
policies, and is dominated by the cost of $2+b$ exponentiations.

There are overheads incurred when storing the ciphertexts in the cloud. First, ABE encryption
introduces overheads that are proportional to the number of attributes per ciphertext. These
overhead are constant in our protocol, because the number of attributes is fixed at $b$.
Second, the data owner may wish to support sliding window policies of different window sizes,
which entails storing multiple ciphertexts for the same data value at the cloud. However, when
encrypting $(k,v)$ for different values of $\beta$ only $E(k,v)$ will be different. As a
result, the overhead is limited to $w.||G||$ where $||G||$ is the number of bits per group element.

\section{Access Control Management}
\label{sec:design} Before starting \trans\ operation, the cloud
checks if the attributes associated with the ciphertext satisfy the
access structure associated with the policy. This is to avoid
performing redundant transformations which return $\bot$. As the
number of streams and policies increase, management and matching of
policies must be done in a systematic and scalable manner. To this
end, we adopt a holistic approach that leverages XACML framework,
allowing the data owner to specify access control policies in XML
format. It also provides a unified framework for managing policies
at the cloud, for defining access control requests and matching them
against the policies. Traditionally, XACML is used for access
control in trusted domains, such as internal systems or in trusted
collaborations. Our work utilizes XACML solely for policy management
and not for access control enforcement. The latter is instead
realized by our cryptographic protocols. 

\subsection{XACML}
XACML is an OASIS framework for specifying and enforcing access control~\cite{xacml}. It
defines standards for writing policies, requests, matching of requests against policies and
interpreting the response. Here, we briefly explain the main components of XACML, more details
can be found in~\cite{xacml}.

Requests in XACML are written in XML, which contain subject credentials and the
system resources being accessed. Subjects and resources are specified in the \emph{Attribute}
elements included in the \emph{Subjects} and \emph{Resources} element respectively. Each policy
in XACML contains a \emph{Target} and a set of \emph{Rules}. A Target element consists of a set
of matching rules that must be met by the request before the rest of the policy can be
evaluated. The XACML framework comprises two main components: a \emph{Policy Enforcement Point}
(PEP) and a \emph{Policy Decision Point} (PDP). Requests are sent to PEP which translates them
to canonical forms before sending to PDP. It interprets responses from PDP and sends the
results back to the users. Policies are loaded into and managed by PDP. After receiving
well-formed requests from PEP, it evaluates them against the loaded policies and sends
the well-formed responses back to the PEP.

\subsection{Using XACML With Our Protocols}
\begin{figure}
\includegraphics[scale=0.55]{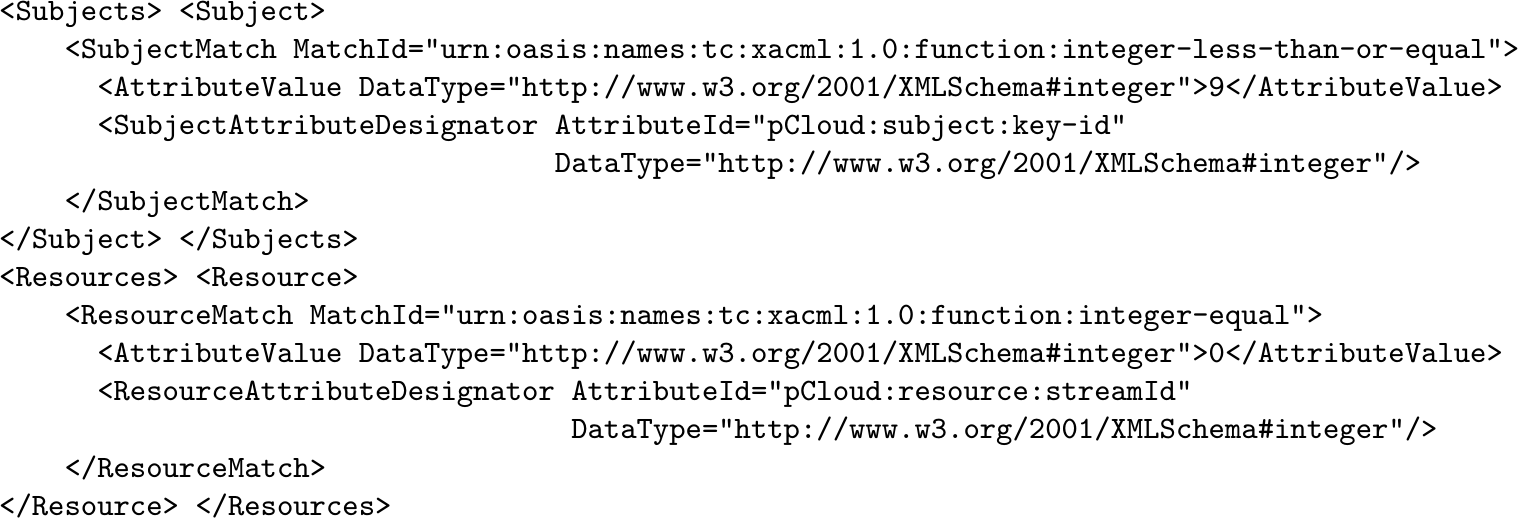}
\caption{Policy for sliding window, $\alpha=9$}
\label{fig:policy}
\end{figure}
\begin{figure}
\includegraphics[scale=0.55]{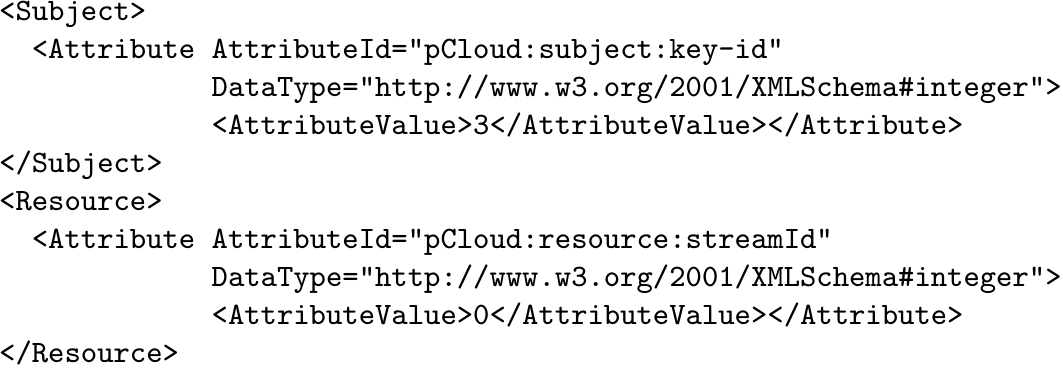}
\caption{Request, $k=3$}
\label{fig:request}
\end{figure}
In traditional settings, resource owners write XACML policies and upload them to a access control
server. Users interested in the resource write XACML requests and submit them to the server,
where they are evaluated by the PDP. If the decisions are Permit, the users are
granted access.

In our setting, the cloud acts as the access control server and runs an XACML instance.
Each data stream is represented as a resource. The data owner writes XACML policies
representing the trigger or sliding window policies. For each policy, the cloud maintains a list
of users to which the policy is given. The user does not have to write or submit XACML
requests to the cloud. Instead, it is the owner who specifies XACML request for every
ciphertext it sends to the cloud. Specifically, the request generated for the tuple $(k,v)$
contains the value of $k$ in the Subject element. Once received at the cloud, it is evaluated
against the loaded policies. The result is a set of matching policies and a set of users to
whom access to the value $v$ should be granted. Figure~\ref{fig:policy} illustrates an XACML policy
for sliding window policies where $\alpha=9$. Figure~\ref{fig:request} shows a request
constructed for the tuple $(k,v)$ where $k = 3$. When evaluated, the PDP will return a Deny
decision, and hence the cloud will not perform transformation for the ciphertext of this tuple.

\section{Prototype and Evaluation}
\label{sec:evaluation}
\subsection{Prototype Implementation}
We implement a prototype\footnote{The source code is available at
\url{https://code.google.com/p/streamcloud/}} demonstrating the
feasibility of the proposed approach and to benchmark the various
components under simple settings. We extend the KP-ABE
implementation from~\cite{libcelia}, which uses PBC
library~\cite{pbc} for pairing operations. Our extensions mainly
concern sliding window policies, which include functions for the
\trans\ operation and for computing the encrypted sums at the cloud.
We use a Java implementation of XACML~\cite{sunxacml} for managing
and matching of policies. The communication between data owners, the
cloud and users is implemented in Java, using sockets and
one-thread-per-connection concurrency model.

\subsection{Microbenchmark}
We set up simple experiments to investigate the costs of the main operations in our protocols.
For each operation, we use latency as a metric for measuring computation cost, which is also
representative of the overhead caused by the operation.

We use the fastest pairing implementation for our experiments (type A~\cite{pbc}), whose base field size is
$512$ bits. There is one data owner sending one stream of data with a certain rate. There is
one data user, to whom the owner specifies either a trigger policy or a sliding policy
(but not both). We fix the sliding window size to $5$ and the maximum data value to $1000$,
which implies that the maximum sum of a window is $5000$. For trigger policies, we set $op$ to be
equality comparison.  Our experiments are run on 3 machines, two desktop machines running as
the owner and the user, and one laptop running as the cloud. Each desktop has 2 2.66Ghz DuoCore
CPUs and 4GB of RAM running Ubuntu Linux. The laptop has 2.3Ghz Core i5 CPU, 4GB of RAM and run
Snow Leopard. All machines are connected via university LAN network.

\begin{figure*}
    \centering
    \subfloat[Data owner]{\includegraphics[scale=0.22, angle=-90]{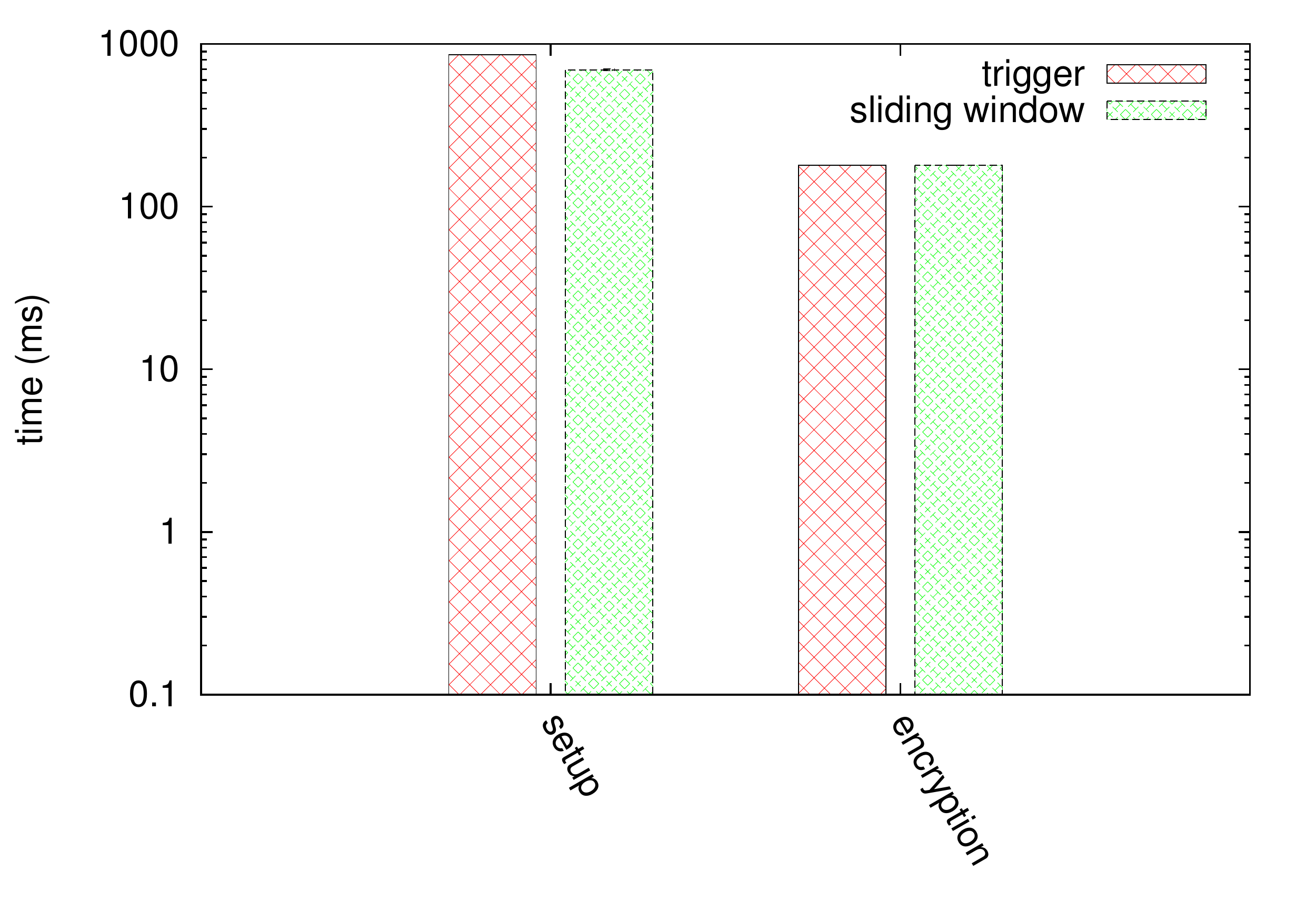}}
    \subfloat[Cloud]{\includegraphics[scale=0.22, angle=-90]{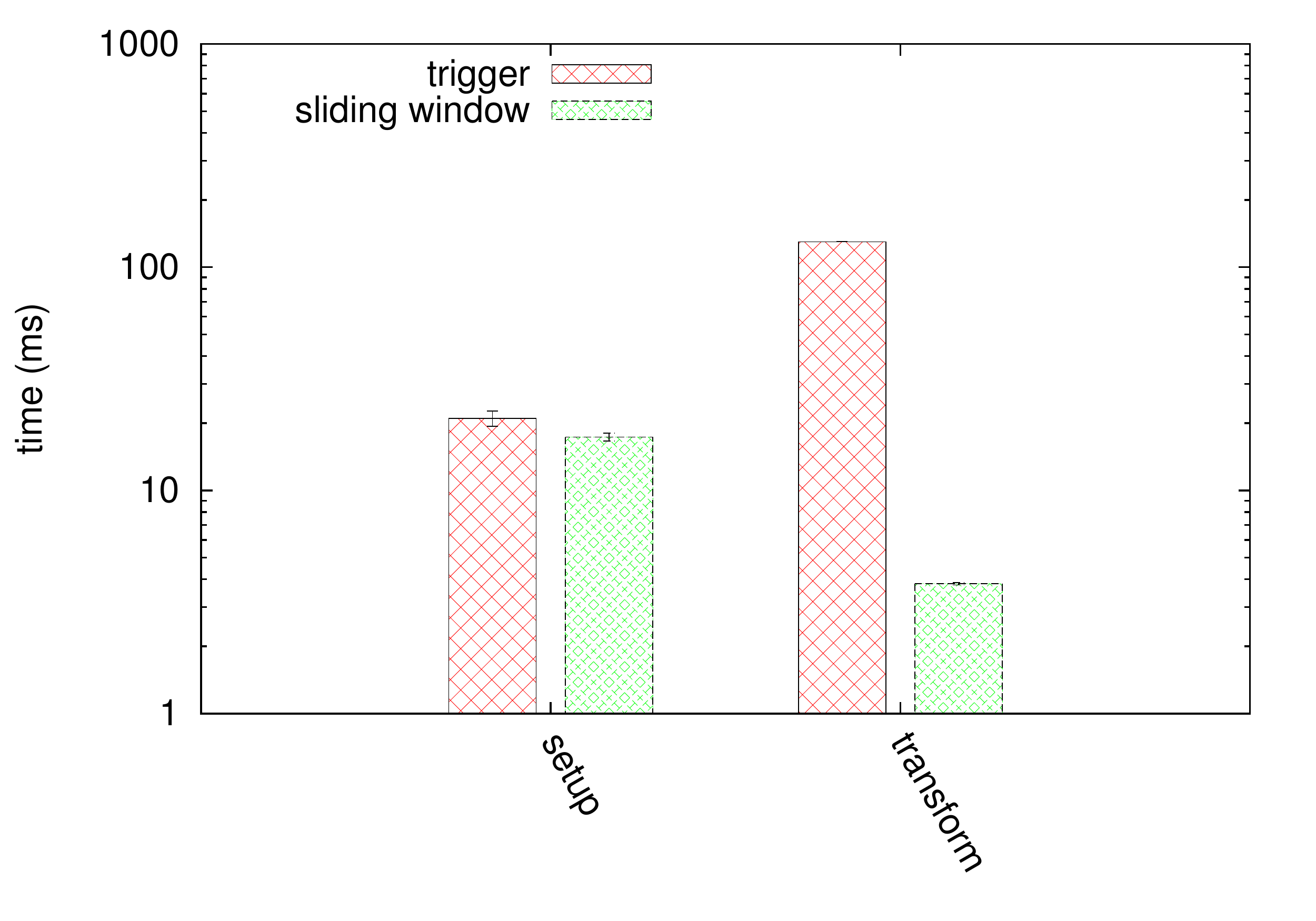}}
    \subfloat[Data user]{\includegraphics[scale=0.22, angle=-90]{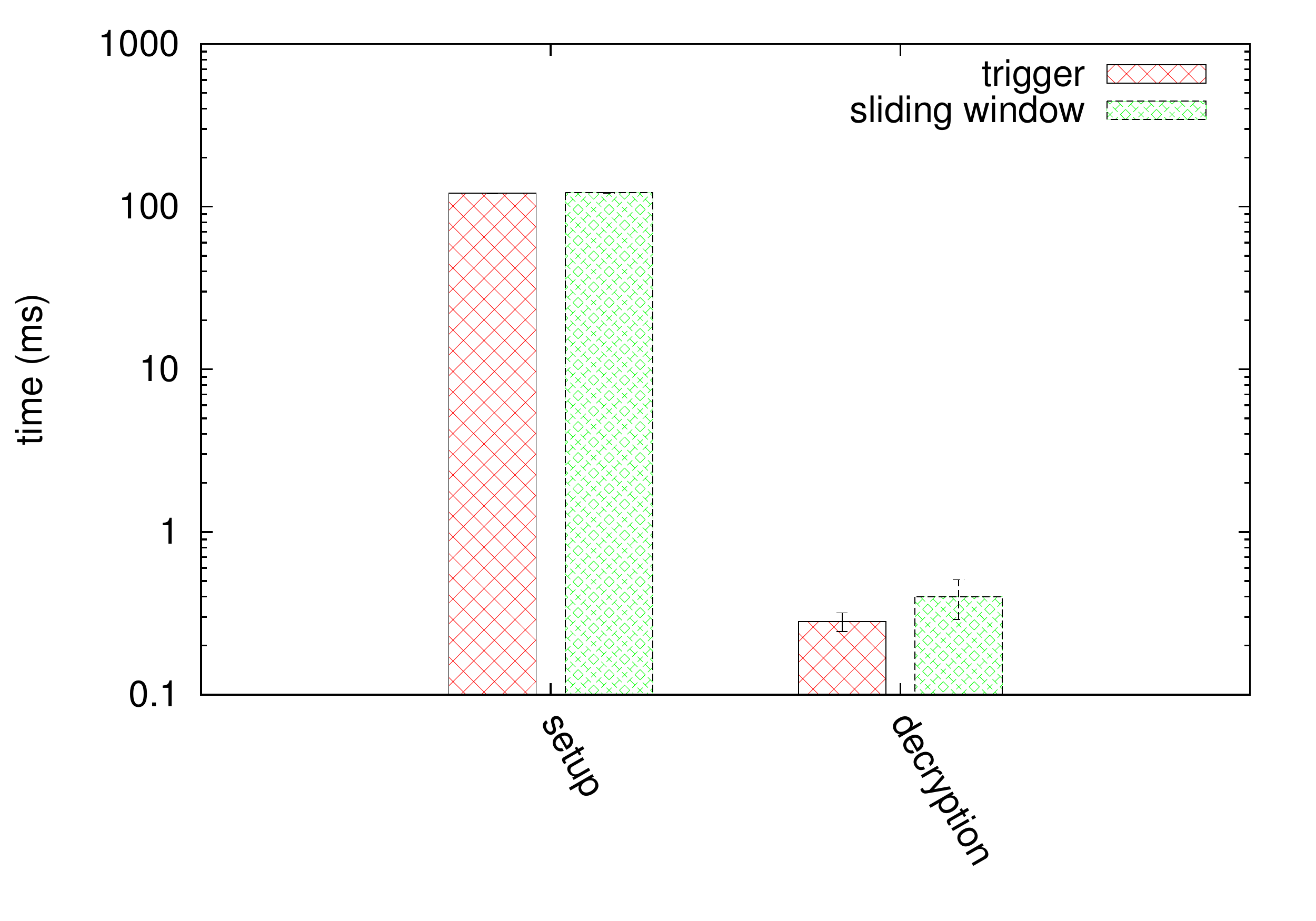}}
    \caption{Time spent on cryptographic operations}
    \label{fig:cryptoTime}
\end{figure*}
Figure~\ref{fig:cryptoTime} depicts the costs of cryptographic
operations. The setup cost for data owner is the highest, since it
involves generating public parameters and transformation key. This
cost dominates the setup time at the user. Recall that this is a one
time cost, and hence does not affect run time performance or
practicality of the approach. The cost for sliding window policies
is slightly smaller than for trigger policies, because the
transformation keys in the former are of smaller sizes (as discussed
in the previous section). Another significant cost in the setup
phase is the cost of pre-computing the look-up table for discrete
logarithms of values in $[0,5000]$. The setup time at the cloud is
much smaller, because it only needs to read and initialize the
public and transformation keys from byte arrays.

The cost per encryption is the same for both sliding window and
trigger policies (as predicted in the previous section), which is
around $179(ms)$. We believe this cost is reasonable, especially
considering that data streams in reality often generate data in very
long intervals (in the order of seconds or even minutes). The
transformation cost at the cloud for trigger policies is orders of
magnitude larger than for sliding window policies. This is because
transformation for the policy $k = \theta$ requires $64$ pairings,
whereas for policy $k \geq \alpha$ only $1$ pairing is needed for
large values of $k$. The maximum latency is around $120(ms)$, which
we also consider as reasonable. Decryption cost at the user is an
order of magnitude smaller than transformation at the cloud, which
further illustrates the benefit of outsourcing bulk of the
computation to the cloud.

\begin{figure}
\centering
\includegraphics[scale=0.23,angle=-90]{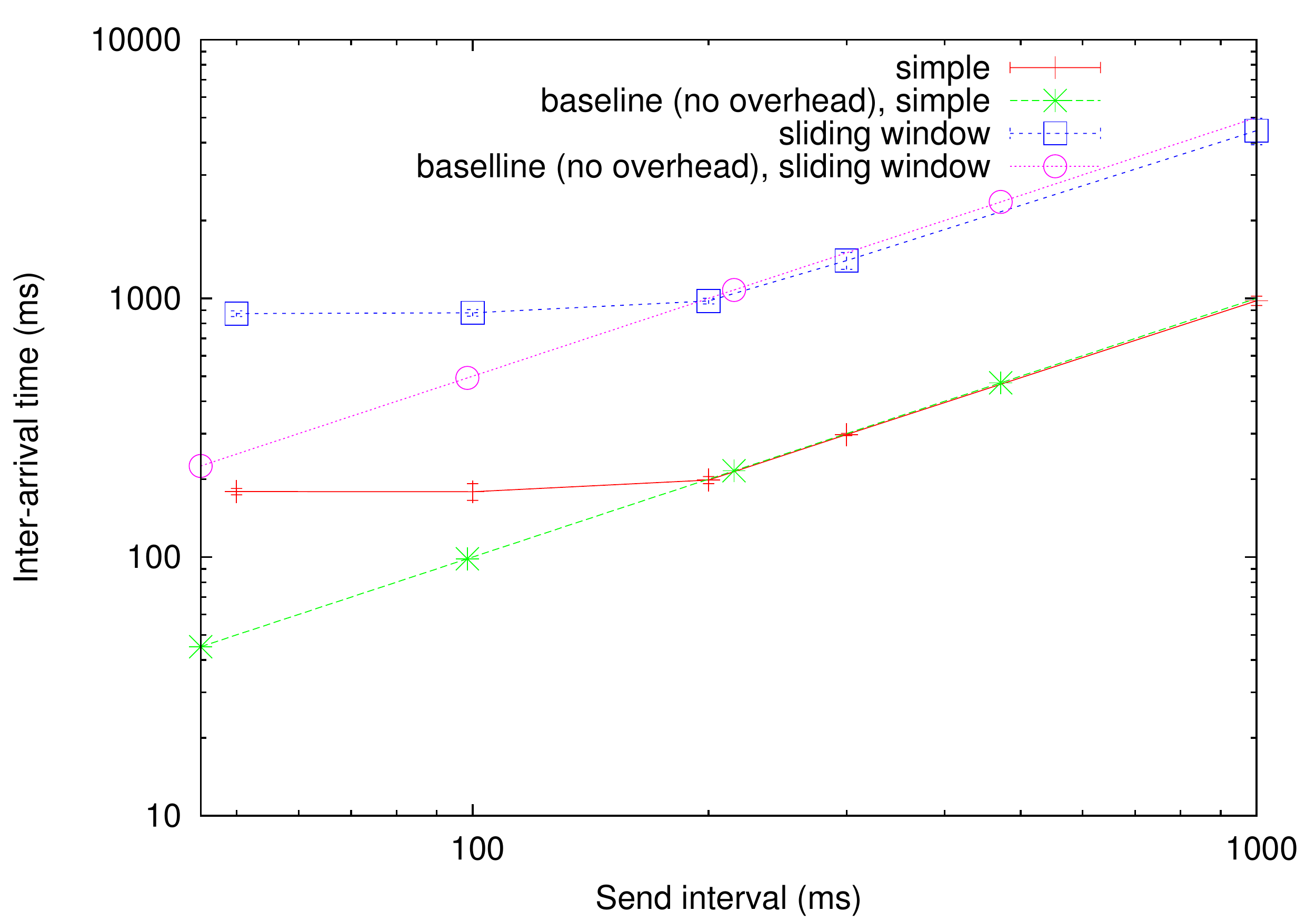}
\caption{Overhead when increasing the sending rate}
\label{fig:rateTime}
\end{figure}
In order to understand the overhead of encryption with respect to
the user-perceived latency, we measure the elapsed time between
decrypted values at the user. For trigger policies, we fix $k$ to be
the same as $\theta$, so that the the user has access to all the
data. Figure~\ref{fig:rateTime} shows the inter-arrival time at the
user. There exists a lower-bound on this metric when we increase the
sending rates (by decreasing the send intervals). This boundary is
exactly at the cost per encryption. This is expected since
encryption becomes the bottleneck at the data owner for high sending
rates.
\begin{figure}
\centering
\includegraphics[scale=0.23,angle=-90]{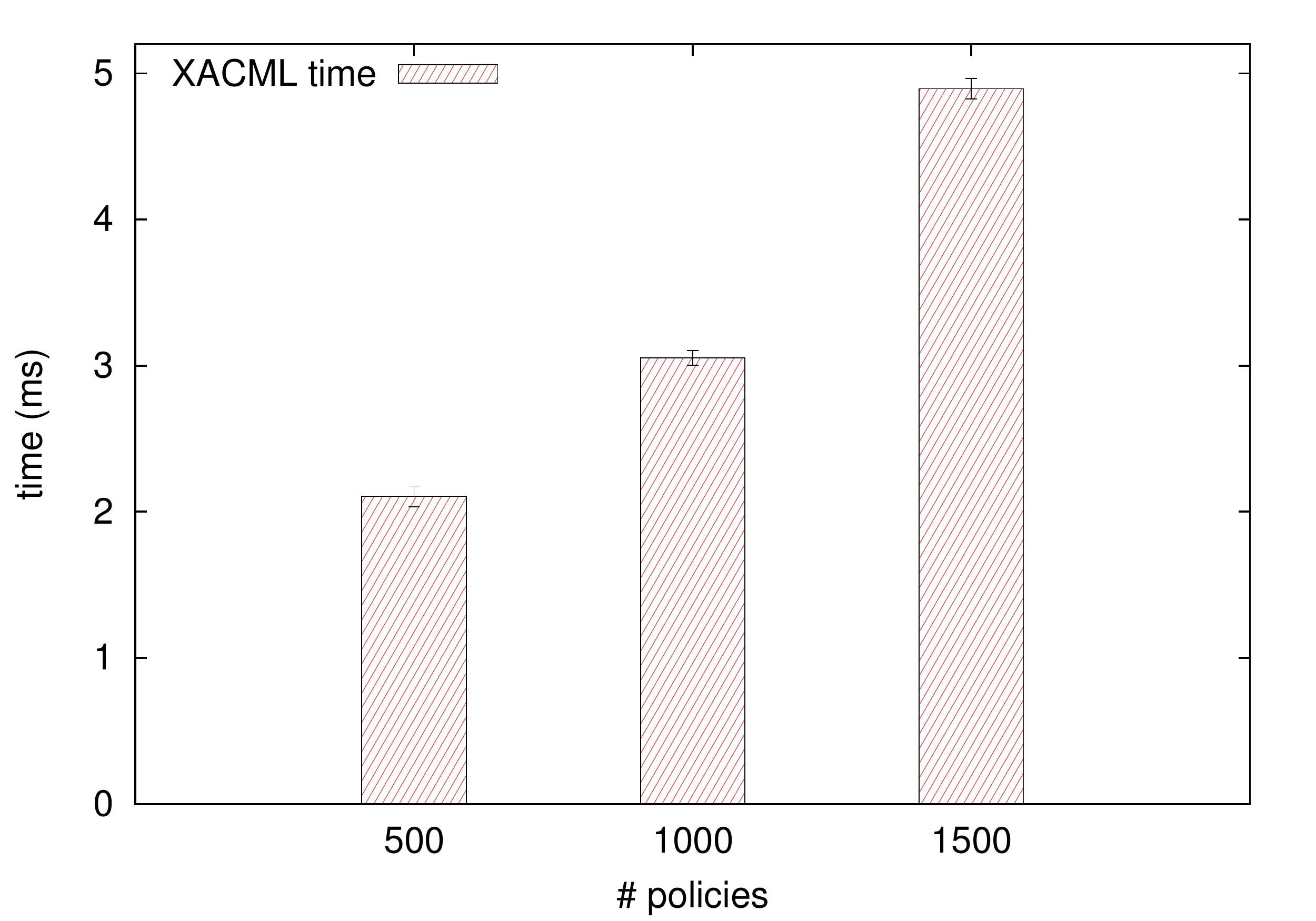}
\caption{XACML processing time}
\label{fig:xacmlTime}
\end{figure}
Finally, we investigate the cost of running the XACML framework. We
upload a large number of policies and send matching requests to the
cloud. The cost for policy matching is depicted in
Figure~\ref{fig:xacmlTime}, which is small and scale gracefully with
more policies. It suggests the system's scalability for large
numbers of policies, streams and users.

\section{Related Work}
\label{sec:relatedWork}
Existing works on access control for stream data assume trusted domains and focus on the
specification and enforcement of access policies~\cite{carminati07,carminati07a,nehme08}. They
are different to our work which considers outsourcing access control to an
untrusted environment. Similarly, previous works that use XACML for fine-grained access control have
also focused on trusted domains~\cite{dinh12}. We use XACML only for policy management, and rely
on encryption schemes for policy enforcement.

When the cloud is untrusted, systems such as CryptDb~\cite{popa11a} and LLP~\cite{zhong07}
proposed systems that guarantee data privacy while still enabling meaningful, database-related
queries on ciphertexts. Our work also provides data privacy, i.e. the cloud cannot learn
plaintext values, but it focuses on the question of who can access the data and what the access
granularity is. Access control on untrusted environment have been investigated
in~\cite{kallahalla03,yu10}, both of which deal with archival data. Our use of ABE allows for more
flexible, fine-grained access control than in~\cite{kallahalla03}. The system proposed
in~\cite{yu10} also employs KP-ABE and the cloud as a
proxy, but it does not work with stream data, nor does it support sliding
window policies. Furthermore, the cloud's main role in~\cite{yu10} is during
key revocation requiring data re-encryption, and unlike in our work, the cloud does not
alleviate the users from expensive computations.

Attribute-Based Encryption schemes are being actively
researched~\cite{ostrovsky07,yu10,goyal06,bethencourt07,lewko11}. Our protocols can be viewed
as an extension of the proxy-based ABE scheme~\cite{green11}. Compared to the original work, we have
built a system that support meaningful access control policies for stream data. We have
additionally integrated XACML
framework for access control management. Finally, we have provided a holistic prototype implementation
and preliminary evaluation for the encryption scheme as well as for the system as a whole.

\section{Conclusions and Future Work}
\label{sec:conclusions}
In this paper, we have proposed a system that allows secure and scalable outsourcing of
access control to the cloud. Our system works with stream data for which access control is more
complex than archival data. We have extended an Attribute-Based Encryption
scheme to support trigger and sliding window policies. We have employed the cloud as a
computational proxy which performs the expensive cryptographic operations on behalf of the user.
In particular, the cloud transforms complex ABE ciphertexts into Elgamal-like ciphertexts that
are decrypted only by authorized users. We have integrated the standard XACML framework to aid policy
management. Thus, our approach for access control outsourcing is holistic and is able to
scale for large numbers of data streams and policies. Preliminary evaluation using a prototype indicates not only the feasibility but also the benefits of access control outsourcing. 

While we have presented a working system, showing a proof-of-concept that outsourcing access
control is feasible and practical for stream data, there is much left for future work. First,
our current prototype is not capable of dealing with large, concurrent workloads. Our immediate
plan is to either improve our prototype using event driven architecture
(SEDA~\cite{seda}) as a replacement for our current one-thread-per-connection concurrency
model, or to use existing high-throughput stream processing engines such as
STREAM~\cite{stream} or Borealis~\cite{cherniack03} followed by experiments at larger scale (more streams, more policies, more users, more cloud servers) over a real cloud infrastructure.

The current system can be extended to support multi-value streams by performing different
encryptions for different values. We have not considered access revocation, which in our case
requires the data owner to change the attribute set during encryption. We plan to investigate if existing
revocable KP-ABE schemes~\cite{attrapadung08} can be integrated in our work, especially if
they allows revocation to be outsourced to an untrusted cloud. Other interesting extensions are to
add support for policies with negative attributes~\cite{ostrovsky07}, and for encryption with
hidden attributes. The former allows for a wider range of access policies, whereas the latter
provides attribute privacy which is necessary when encryption attributes are the actual data.

Finally, we would like to relax the trust assumption for the cloud. Currently, the cloud
performs all the computations delegated to it in an honest manner. However, a malicious cloud
which skips or distorts computations will greatly increase the cost of security while maintaining a good service quality.
\bibliographystyle{plain}
\bibliography{abeAC,paper}
\end{document}